\documentclass{ws-p8-50x6-00}
\usepackage{amssymb}
\begin{document}
\def\lsim{\:\raisebox{-0.5ex}{$\stackrel{\textstyle<}{\sim}$}\:}
\def\gsim{\:\raisebox{-0.5ex}{$\stackrel{\textstyle>}{\sim}$}\:}
\def\JHEP{\em Journal of High Energy Physics}
\def\EPJ{\em Eur. Phys. J. }
\def\sh{mbox{\cal h}}
\def\et{\mbox{${E\!\!\!/}_T$}} 
\def\rpv{\mbox{${R\!\!\!/}_p$}} 
\def\bv{\mbox{${B\!\!\!/}$}} 
\def\lv{\mbox{${L\!\!\!/}$}} 
\def\rp{\mbox{$R_p$}}
\def\eplem{\mbox{$e^+e^-$}}
\def\emem{\mbox{$e^-e^-$}}
\def\game{\mbox{$\gamma e$}}
\def\gamgam{\mbox{$\gamma \gamma$}}
\def\ttbar{\mbox{$\bar t t$}}
\def\tldch{\mbox{$\tilde{\chi}$}}
\def\stau1{\mbox{$\tilde \tau_1$}}
\def\N0{\widetilde \chi^0}
\def\Cp{\widetilde \chi^+}
\def\Cm{\widetilde \chi^-}
\def\Cpm{\widetilde \chi^\pm}
\def\Cmp{\widetilde \chi^\mp}
\def\mET{E_T \hspace{-1.1em}/\;\:}
\def\mpT{p_T \hspace{-1em}/\;\:}
\title{SUSY at the Linear Colliders: Working Group Summary}

\author{Rohini M. Godbole}

\address{Center for Theoretical Studies, Indian Institute of Science,
Bangalore 560 012,
INDIA\\E-mail: rohini@cts.iisc.ernet.in}


\begin{flushright}
IISc-CTS/19/00 \\
hep-ph/0011236 
\end{flushright}

\vskip 25pt
\begin{center}

{\large\bf SUSY at the Linear Colliders: Working Group Summary
\footnote{Invited talk presented at the 
III ACFA Linear Collider Workshop, August 2000, Taipei, Taiwan.}}    
       \\
\vskip 25pt

{\bf                        Rohini M. Godbole } \\ 

{\footnotesize\rm 
                      Centre for Theoretical Studies, 
                     Indian Institute of Science, Bangalore 560 012, India. \\ 
                     E-mail: rohini@cts.iisc.ernet.in  } \\ 

\vskip 20pt

{\bf                             Abstract 
}

\end{center}
\begin{quotation}
I summarise the activities of the different members of the
SUSY working group. There have been two major areas of activity:
1) precision measurement of the SUSY particle masses/couplings and hence those
of the SUSY model parameters, 2) investigations into
SUSY searches at \eplem , \gamgam , $\game$ and $\emem$ colliders, in the 
nonstandard scenarios such as explicit CP violation, R-parity violation  
and Anomaly Mediated Supersymmtery Breaking. In addition there have been
studies which looked at  the effect of  `large' extra dimensions at the 
various colliders mentioned above.
\end{quotation}
\maketitle

\abstracts{ 
I summarise the activities of the different members of the
SUSY working group. There have been two major areas of activity:
1) precision measurement of the SUSY particle masses/couplings and hence those
of the SUSY model parameters, 2) investigations into
SUSY searches at \eplem , \gamgam , $\game$ and $\emem$ colliders, in the 
nonstandard scenarios such as explicit CP violation, R-parity violation  
and Anomaly Mediated Supersymmtery Breaking. In addition there have been
studies which looked at  the effect of  `large' extra dimensions at the 
various colliders mentioned above.
}

\section{Introduction}
In this talk I would like to summarise some of the investigations carried
out in the context of Supersymmetry as well as related ideas of `large' 
extra dimensions in the Asian Supersymmetry Subgroup.
For \eplem\ colliders, some of the studies concerned  the precision measurements of 
sparticle masses, cross-sections  and consequent extraction 
of the parameters of the SUSY model\cite{1,2,3} in the context of
(C)MSSM whereas some looked at somewhat nonstandard aspects such as
effects of explicit CP violation on the $\tilde t$ phenomenology\cite{4}
or effect of \rpv\ on the sparticle searches\cite{5} as well as
some explicit search strategies for the scenario with Anomaly Mediated SUSY 
Breaking(AMSB)\cite{6}.  A theoretical study of trying to extract 
CP phases of the SUSY model in a study of correlated production and decay 
of neutralinos\cite{14} has also been done. 
Effects of \rpv\  at \gamgam\cite{7}, \game\cite{8}
and \emem\cite{9} colliders have been looked at in a series of invsetigations.
Possibilities of looking for the effect of the `large' extra dimensions
at the \eplem\ colliders via the phenomenology of radions\cite{10}, 
at the \gamgam\ colliders via the dijet/$\ttbar$  production\cite{11,11p} and 
finally at \game\cite{12} as well as  \emem\cite{13} colliders via 
graviton production or via indirect effects respectively, have also been 
investigated.  While it is not possible to give details of all these,
I will choose a few and discuss those results in some detail.

\newpage
The list of various topics and the associated investigators is first given
below:
\begin{enumerate}
 \item{\eplem\ colliders}
  \begin{description}
   \item[1] Reconstruction of chargino system at \eplem\ colliders with(out)
polarisation\cite{1} : S.Y. Choi, A. Djouadi, H. Dreiner, M. Guchait, 
J. Kalinowski, H. Song and P. M. Zerwas
   \item[2] Study of $\eplem\ \to \tldch_{1}^{+} \tldch_{1}^{-}
\to {\tilde{\tau}}^{-} {\bar\nu_{\tau}} {\tilde{\tau}}^{+} \nu_{\tau}$ and
determination of $m_{\tldch_{1}^{\pm}}$\cite{2} : Y. Kato, M. Nojiri, K. Fujii, 
T. Kamon
\item[3] Effect of radiative corrections on kinematic reconstruction of the 
squark mass\cite{3} : M. Drees, Oscar J.P. Eboli, R.M. Godbole and S. Kraml.
\item[4] $\eplem\ \to \tilde {t_{1}} {\tilde{t}}_{1}^{*} h$ with
 explicit CP violation\cite{4} : S. Bae
\item[5] $\eplem\ \to \tldch_{i}^{+} \tldch_{j}^{-} ,
\tldch_{i}^{0} \tldch_{j}^{0}$ and cascade decays in \rpv\ theories\cite{5}
: D. Ghosh, R.M. Godbole and S. Raychaudhuri
\item[6] Linear collider signals of a Wino LSP in AMSB\cite{6} : D. Ghosh,
P. Roy, S. Roy
\item[7] CP phases in correlated production and decay of
$\tldch_{j}^{0}$ in MSSM with explicit CP violation\cite{14}:  S.Y. Choi, H.S. Song 
and W.Y. Song.
\end{description}
\item \gamgam\ colliders
 \begin{description}
  \item[1] Sfermion production and decay through \rpv\ interactions\cite{7}: 
D.  Choudhury, A. Datta
 \end{description}
 \item $e\gamma$ colliders
  \begin{description}
  \item[1] $e\gamma \to \tilde{e} \tldch_{1}^{0}$ followed by \rpv\ decay
of $\tldch_{1}^{0}$~\cite{8} : D. Ghosh, S. Raychaudhuri
  \end{description}
 \item \emem\ colliders
   \begin{description}
    \item[1] Signals for \rpv\ at \emem\ colliders\cite{9} : D. Ghosh, S. Roy
  \end{description}
  \item `Large' extra dimensions
   \begin{description}
    \item[1] Phenomenology of a radion in Randall-Sundrum Scenario at
colliders\cite{10} : S. Bae, P. Ko, A.S. Lee, J. Lee
    \item[2] `Large' extra dimensions and dijet production\cite{11} : 
D.Ghosh, P. Mathews, P.  Poulose and K. Sridhar. 
    \item[3] `Large' extra dimensions and \ttbar\ production\cite{11p} : 
P. Mathews, P.  Poulose, K. Sridhar
    \item[4] $e\gamma$ colliders and TeV scale Quantum Gravity\cite{12} : 
D. Ghosh, P. Poulose and K. Sridhar.
    \item[5] Randall-Sundrum Model and \emem\ colliders\cite{13} : 
D. Ghosh, S.  Raychaudhuri
 \end{description}
\end{enumerate}

\section {Precision SUSY studies and investigations into \rpv\ , CP violation
 in SUSY}
\subsection{Reconstruction of chargino system and hence SUSY model
parameters.\protect\cite{1}}
\begin{figure}[htb]
       \centerline{
      \includegraphics*[scale=0.365]{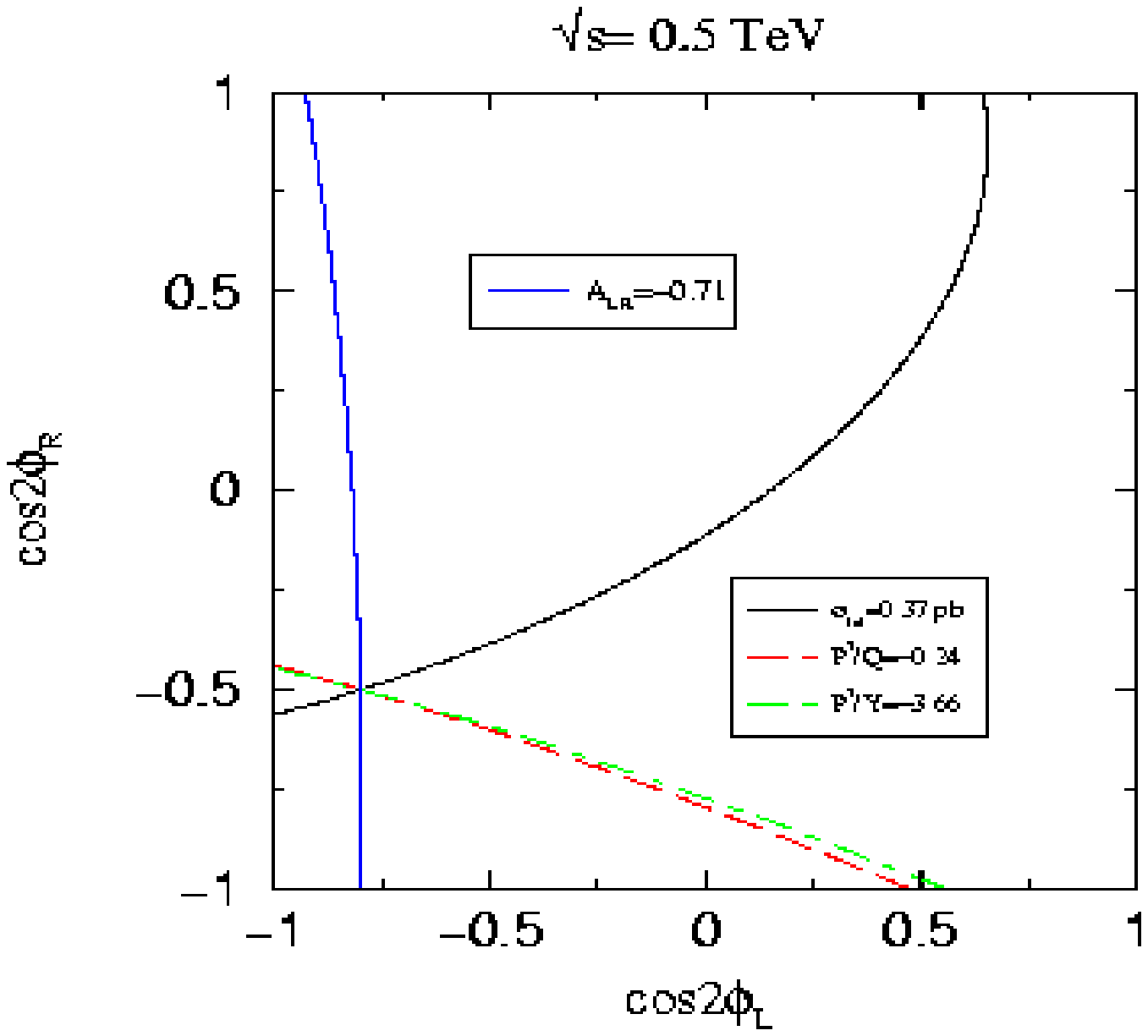}
      \includegraphics*[scale=0.35]{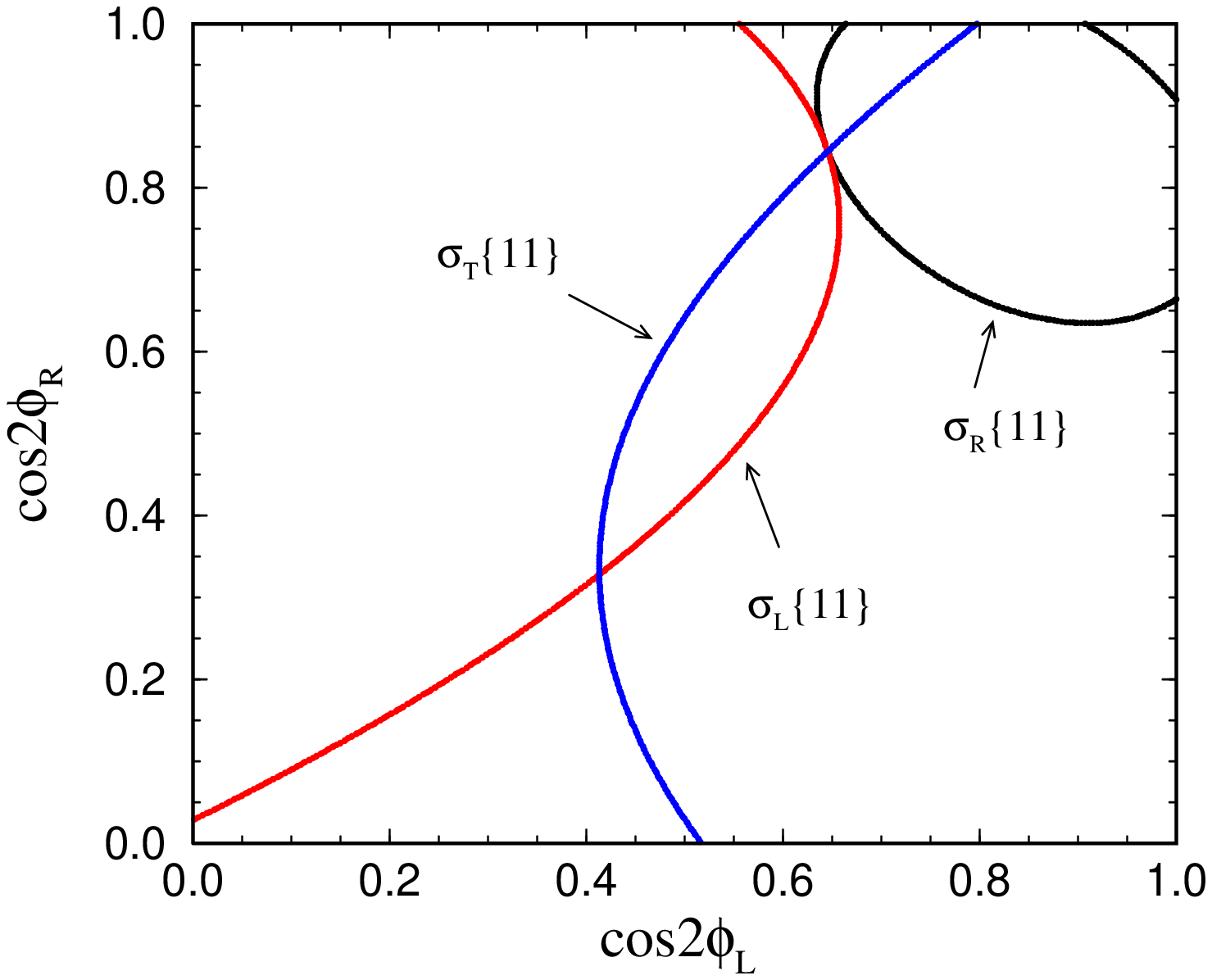}
  }
\caption{\em Use of initial or final state polarisation  in the 
determination of $\cos 2\phi_L$ and $ \cos  2 \phi_R$.\protect\cite{1}}
\label{fig1}
\end{figure}

These authors have shown how   a study of the process
$\eplem\ \to \tldch_{i}^{+} \tldch_{j}^{-}$ allows for extraction of
all the model parameters relevant for  the chargino sector 
in the (C)MSSM.  $\tldch_{1L}, \tldch_{1R}$ can be written in terms of the
gaugino-higgsino basis as
$$
\tldch_{1L} = \widetilde {W_{L}} \; 
cos\phi_{L} + \widetilde {H_{1L}} \; sin\phi_{L} ; \;\;\;\;
\tldch_{1R} = \widetilde {W_{R}} \; 
cos\phi_{R} + \widetilde{H_{2R}} \; sin\phi_{R}
$$
The mass $m_{\tldch_{i}}$ can be determined from the sharp threshold
rise. The other variables to be measured are $\sigma_{tot}$, chargino
polarisation and spin-spin correlation. The important part of the work is a
method to reconstruct the last, through angular distribution of decay products.
They identify functions ${\cal P}$, ${\cal Q}$, ${\cal Y}$ such that 
${\cal P}^{2}/{\cal Q}$, ${\cal P}^{2}/{\cal Y}$ can be obtained from all the
observed variables and dynamics dependent quantities cancel from the ratio.
Here one uses unpolarised beam but makes use of the polarisation of the
produced chargino. The contours of constant $\sigma^{tot}$,
${\cal P}^{2}/{\cal Q}$ and ${\cal P}^{2}/{\cal Y}$ intersect at one point in
$\cos 2\phi_{L}$ - $\cos 2 \phi_{R}$ plane as shown in the left panel of
Fig.~\ref{fig1}.
The knowledge  of $m_{\chi_{1}^{\pm}}$ along with
$\cos 2\phi_{L},\cos 2\phi_{R}$ then allows one to determine the 
parameters $M_2$, $\mu$ and $tan\beta$ upto a two fold ambiguity.
Instead of using the polarisation information of the produced charginos, if
polarised beams  are available, measurement
of $\sigma_{L}^{11}, \sigma_{R}^{11}$
(where the superscripts 1,1 stand for the $\tldch_{1}^{+}
\tldch_{1}^{-}$ production), allows determination of  the mixing angles 
$cos2\phi_{L}, cos2\phi_{R}$ upto a four fold ambiguity, assuming that one has 
the knowledge of $m_{\tldch_{1}^{+}}$,  $m_{\tilde {\nu}}$ from kinematics, 
and further assuming $g_{\tilde {\nu} \tilde {W} l} = g_{2}$. 
Use of cross-sections $\sigma_{T}^{11}$ with transeverse polarisation helps
remove this ambiguity as shown in the right panel of Fig.~\ref{fig1}.
If production of all the charginos is allowed kinematically, unambiguous
reconstruction of $\mu, M_{2}$ and tan$\beta$ is possible  even without 
resorting to the use of transverse polarisation or the complicated study of
the polarisation of the produced charginos. One can even test
the relation $g_{e \tilde {\nu} \widetilde {W}} = g_{e \nu W}$.
The results of an analysis, including only statistical errors, 
for $\int {\cal L} dt = 1ab^{-1}$ are shown  in Table~\ref{tab:tab1}.
The left (right) column shows the input (extracted) values for the 
two chargino masses and $\cos 2 \phi_{L}, \cos 2 \phi_{R}$.
\begin{table}[htb]
\caption{\em Comparison of the input and output values of the mixing angles in 
the chargino sector extracted from different  chargino measurements for
$\int {\cal L} dt = 1 {\rm ab}^{-1}$.\label{tab:tab1}}
\begin{center}
\begin{tabular}{|c|c|}
\hline
 Input  &  Extracted \\ \hline
&\\
$m_{\tldch_{1}^{\pm}}$ = 128 GeV,&
$m_{\tldch_{1}^{\pm}}$ = 128 $\pm 0.04 $GeV, \\
$m_{\tldch_{2}^{\pm}}$ = 346 GeV.& $m_{\tldch_{2}^{\pm}} = 346$
$\pm 0.25$ GeV. \\ 
&\\\hline
&\\
$\cos 2 \phi_{L} = 0.645$,& $\cos 2 \phi_{L} = 0.645 \pm 0.02,$\\
$\cos 2 \phi_{R} = 0.844.$ & $\cos 2 \phi_{R} = 0.844 \pm 0.005 $. \\ 
&\\\hline
&\\
$g_{e \tilde {\nu} \tilde {W}} / g_{e \nu W}  =  1$ &
$g_{e \tilde {\nu} \tilde {W}} / g_{e \nu W}  =  1 \pm 0.01 $ \\ 
&\\ \hline
\end{tabular}
\end{center}
\end{table}
Using these uniquely determined values of $\cos 2 \phi_{L}, \cos 2\phi_{R}$ and
the chargino masses, one can then extract  $M_{2}$, $\mu$ and tan$\beta$ uniquely.
\begin{table}[t]
\caption{\em The input and output values of $M_2,\mu $ and 
$\tan\beta$ for two input points for  an integrated luminosity 
$1 ab^{-1}$.\label{tab:tab2}}
\begin{center}
 \begin{tabular}{|c|c|c|c|c|}
\hline
 parameter & Input & Extracted & Input & Extracted \\ \hline
$M_{2}$ & 152 & $152 \pm 1.75$ & 150 & $150 \pm 1.2 $ \\ \hline
$\mu$ & 316 & $316 \pm 0.87 $ & 263 & $ 263 \pm 0.7 $ \\ \hline
$tan \beta$ & 3 &  3 $\pm$ 0.69 & 30 & $ > $ 20.2 \\ \hline
 \end{tabular}
\end{center}
\end{table}
Table~\ref{tab:tab2} shows again the input and extracted values of
$M_2,\mu,\tan\beta$ for two different inputs. We see that the determination 
can be quite precise, except in the situation with large tan$\beta$.
This is easy to understand as all the chargino variables are proportional
to $\cos 2\beta$. 
Recall, however, that the $1\sigma$ errors shown here are purely statistical.
An analysis including full detector effects,  using this beautiful method
which does not even require beam polarisation, might indeed be worthwhile in 
view of its promise.  

\subsection{Determination of $m_{\tldch_{1}^{+}}$ through its
$\tilde {\tau_{1}}$ decay.\protect\cite{2}}
These authors looked at the determination of the chargino mass
$m_{\tldch_{1}^{\pm}}$, in the large tan $\beta$ (tan $\beta > $ 40) case where
one expects a light stau $\tilde{\tau_{1}}$. If a mass heirarchy (expected
at large $tan \beta$ in (M)SUGRA scenario as well) : $m(\tilde{l}) >
m(\tldch_{1}^{+}) > m(\tilde{\tau_{1}}) > m(\tldch_{1}^{0})$ exists, then
$\tldch_{1}^{+}$ decays into a $\tilde{\tau_{1}}\nu_{\tau}$ almost 100 \% of the
time. Thus the process used to study $\tldch_{1}^{+} \tldch_{1}^{-}$
production will now be
$\eplem \to \tldch_{1}^{+} + \tldch_{1}^{-} \to$
${\tilde{\tau}}_{1}^{+} \nu_{\tau} {\tilde{\tau}}_{1}^{-} \bar{\nu_{\tau}} \to$
$\tau^{+} \tau^{-} \tldch_{1}^{0} \tldch_{1}^{0} \nu_{\tau}
\bar{\nu_{\tau}}$.
For the mass heirarchy given above, we can assume $m_{{\tilde{\tau}}_1},
m_{\tldch_{1}^{0}}$ to be known from studies of ${\tilde{\tau}}_{1}^{+}
{\tilde{\tau}}_{1}^{-}$ production. The main backgrounds are 
$2\gamma$, WW and ZZ.  Using the $E_{jet}$ distribution ($\tau$ is 
detected as a thin jet) for $\int {\cal L}
dt = 200 fb^{-1}$, with an input $m_{\tldch_{1}^{+}}$ = 172 GeV,
$m_{\tilde{\tau_{1}}}$ = 152 GeV and $m_{\tldch_{1}^{0}}$ = 87 GeV, the authors
find $m_{\tldch_{1}^{+}} = 171.3 \pm 0.5$ GeV. The corresponding 
$\Delta \chi^2 $ is shown in 
\begin{figure}[htb]
       \centerline{
      \includegraphics*[scale=0.3]{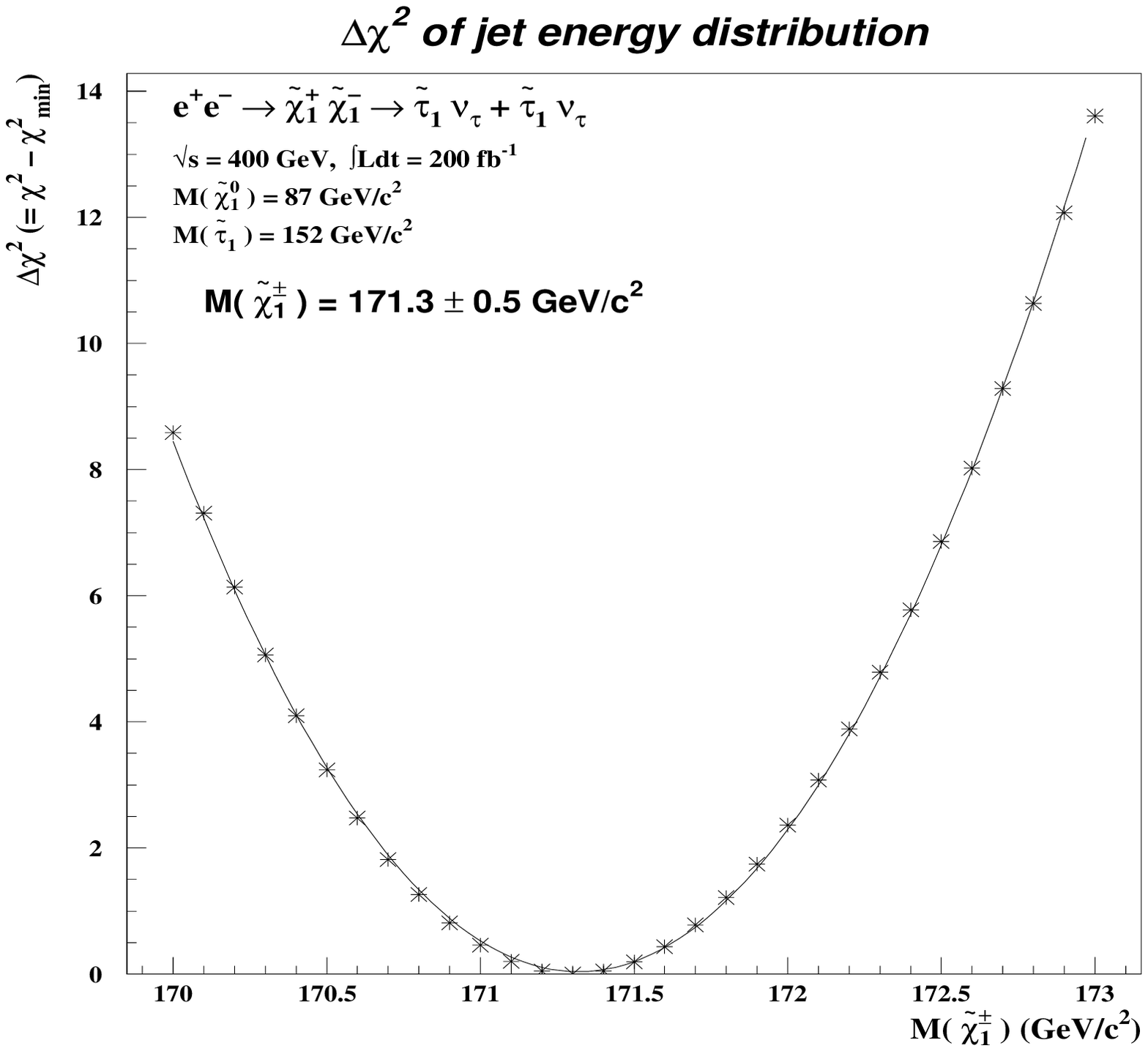}
  }
 \caption{\em  $\Delta \chi^2$  as a function of 
$m_{\tldch_{1}^{+}}$ for an integrated luminosity of 200 fb$^{-1}$.}
\label{fig2}
\end{figure}
Fig.~\ref{fig2}.
\subsection{Effects of radiative corrections on kinematic reconstruction of
squark mass.\protect\cite{3}}
\begin{figure}[htb]
       \centerline{
      \includegraphics*[scale=0.35]{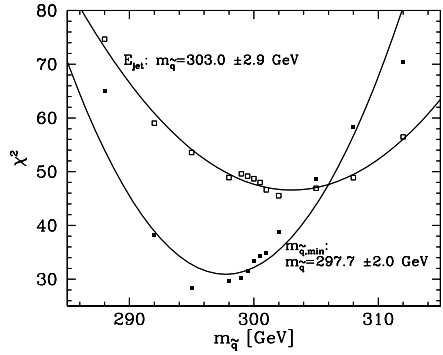}
 } 
\caption{\em $\chi^{2}$ as a function of $m_{\tilde q}$  for an integrated 
luminosity of 50 fb$^{-1}$.\protect\cite{3}}
\label{fig3}
\end{figure}

The authors studied here
 $\eplem\  \to \tilde{q} {\tilde{q}}^{*}$.
They included the radiative corrections to production and decay as well as the
effects of the ISR. 
They used two estimators for the mass of $\tilde{q}$ : 
1) $m_{\tilde{q},min}$\cite{fengfinnell} for two body final states and 2)
$E_{jet}$ distribution. For an integrated luminosity of $50fb^{-1}$ and an input
value of $m_{\tilde{q}}$ = 300 GeV, the authors found $m_{\tilde{q}}$ = 297.7 $\pm$ 2 GeV
and $m_{\tilde{q}}$ = 303 $\pm$ 2.9 GeV from the two estimators
$m_{\tilde{q},min}$ and $E_{jet}$ distributions respectively. The results are
shown in Fig.~\ref{fig3}.
 Thus it is seen that the effect of higher order corrections to decays does not
deteriorate the utility of the estimator $m_{\tilde{q},min}$. The effect of
hadronisation is not yet included in this analysis.
\subsection{$\eplem \to \tilde{t_{1}} \tilde{t_{1}^{*}}h$ with explicit CP
violation.\protect\cite{4}}
 The author here has looked at explicit CP violation in the MSSM higgs
sector and loop effects (essentially the effects of large third generation
trilinear term) on the Higgs potential with complex $A_{t}$.
Essentially changes in $m_{{\tilde{t}}_{1}}, g_{{\tilde{t}}_{1}
{\tilde{t}}_{1}^{*} h}$ due to loop effects were calculated.
 There are two CP violating phases Arg($\mu$) and Arg($A_{t}$). The author
chooses to satisfy the electric dipole constraints as well as cosmological
ones\cite{pilaftsis} : i) Arg($\mu) < 10^{-2}$  ii) $m_{\tilde{g}} > 400$
GeV  (iii) $ |A_{e}|, |A_{u,c}|, |A_{d,e}| < 10^{-3} |\mu|$ and iv) maximal
mixing in the stop sector $|A_{t}| = |\mu cot\beta |$. 
\begin{figure}[htb]
       \centerline{
      \includegraphics*[scale=0.45]{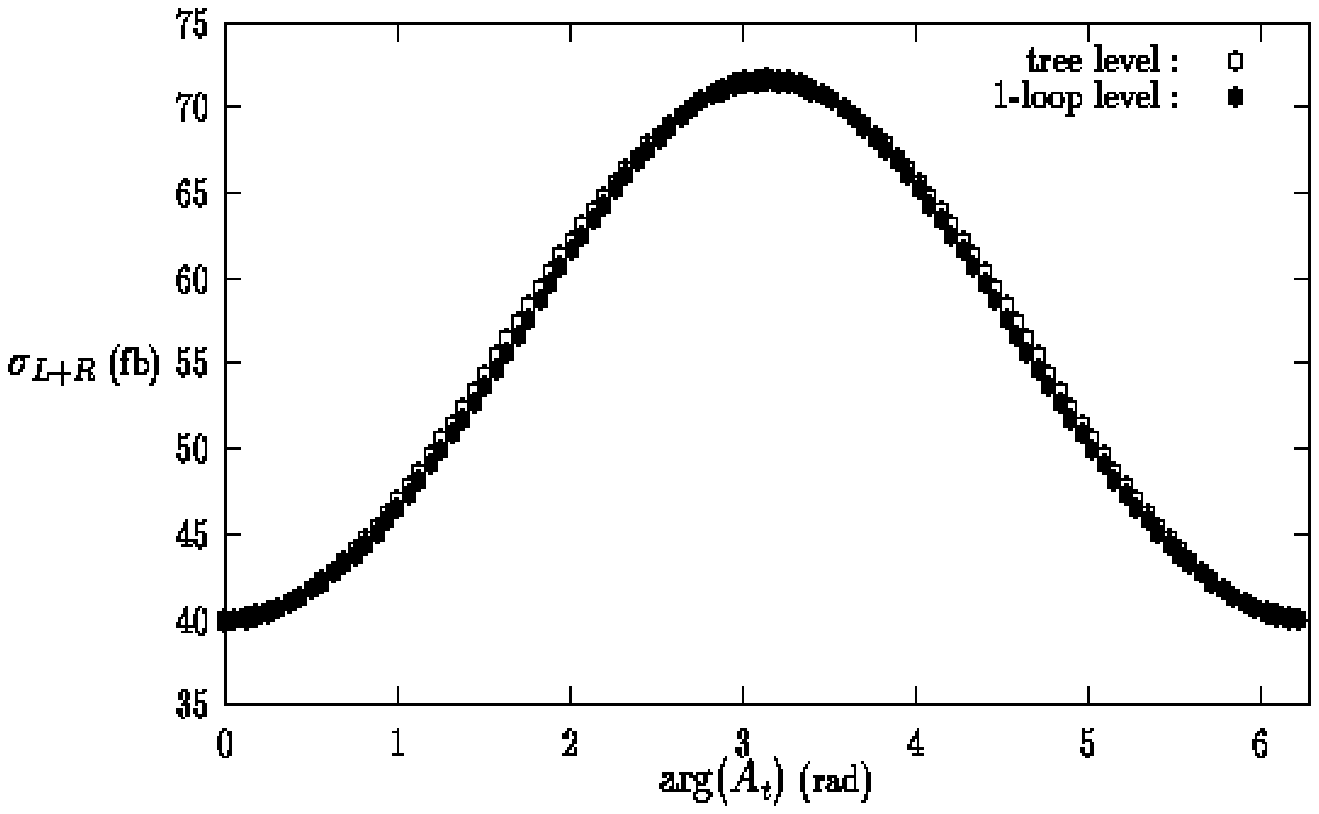}
      \includegraphics*[scale=0.45]{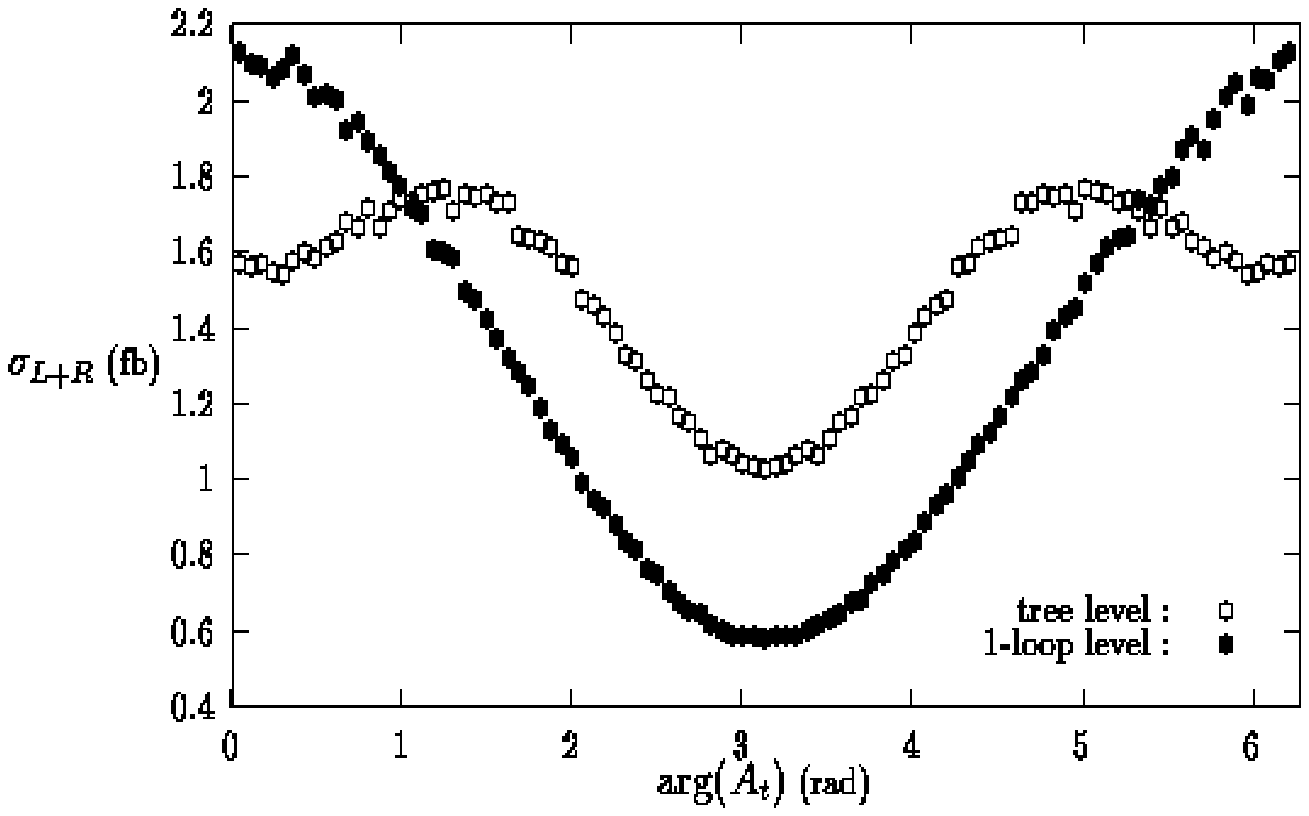}
  }
   \caption{\em  Effect of loop corrections on 
$\sigma(\tilde{t_{1}} \tilde{t_{1}}^{*})$ (left) and 
$\sigma(\tilde{t_{1}} \tilde{t_{1}}^{*}h)$ (right) as a function of $ Arg(A_t)$ 
($\sqrt{s} = 500$ GeV, $\mu = 500$ GeV, $|A_t| = 250$ GeV, $M_A \simeq  194$ 
GeV, $\tan \beta = 2$, and $M_{\rm SUSY} = 500 $ GeV).}
\label{fig4}
\end{figure}
Fig.~\ref{fig4} shows $\sigma_{L+R}(\eplem \to {\tilde{t}}_{1}
{\tilde{t}}_{1}^{*})$ and $\sigma_{L+R}(\eplem \to {\tilde{t}}_{1}
{\tilde{t}}_{1}^{*}h)$ in the left and the right panels respectively. 
We see that though the loop effects are minimal in ${\tilde{t}}_{1}
{\tilde{t}}_{1}^{*}$ production, the dependence on arg($A_{t}$) is quite
strong; on the other hand for $\eplem \to {\tilde{t}}_{1}
{\tilde{t}}_{1}^{*}h$ the cross sections are quite small but loop effects
are substantial and can be as much as 100 \%. The figure shows this for values
of parameters given in the figure caption. The interesting cross-sections
are $\sim$ few fb.
A possible discussion of extracting $|A_{t}|, Arg(A_{t})$ by combining the
measurements of these cross-sections with the knowledge of higgs masses has
been sketched and seems worth pursuing.
\subsection{Chargino/neutralino production and cascade decays of LSP through
\rpv\ couplings at \eplem\ colliders.\protect\cite{5}}
\begin{figure}[htb]
       \centerline{
      \includegraphics*[scale=0.5]{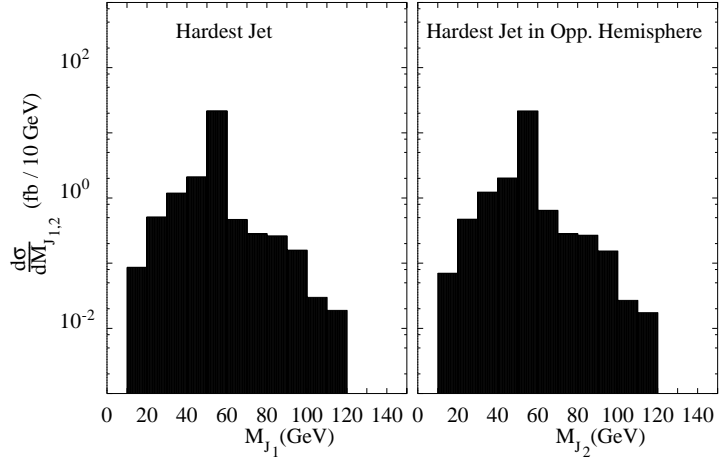}
  }
   \caption{\em  
 Illustrating the distribution in invariant mass
reconstructed from ($a$) the hardest jet and all jets in the same
hemisphere, and ($b$) the hardest jet in the opposite hemisphere and all
remaining hadronic jets, when all contributions are summed over (signal
only), for a chosen point in the parameter space.\protect\cite{5}}
\label{fig5}
\end{figure}

In this work the authors consider $\eplem \to \tldch_{i}^{+} \tldch_{j}^{-}$
and $\eplem \to \tldch_{i}^{0} \tldch_{j}^{0}$. Once the LEP constraints on
$m_{\tldch_{1}^{\pm}}$ are imposed, it is found that over a large part of 
the region of parameter space which allows $\tldch_{1}^{+}$ within the 
reach of a 500 GeV linear collider, $\tldch_{3}^{0}, \tldch_{4}^{0}$ 
and $\tldch_{2}^{+}$ are almost always beyond its reach, at least in
the framework of the (C)MSSM.  Hence it is sufficient to consider 
 i) $\eplem \to \tldch_{1}^{0} \tldch_{2}^{0}$  ,
ii)$\eplem \to \tldch_{2}^{0} \tldch_{2}^{0}$
and iii) $\eplem \to \tldch_{1}^{+} \tldch_{1}^{-}$. Further, using the
approximate degeneracy of $\tldch_{1}^{\pm}$ and $
\tldch_{2}^{0}$, the number of decay chains to be considered are reduced to
managable numbers. The authors work in the weak coupling limit of \rpv\ 
and consider the effect of \rpv\ only for the LSP decay.
For the $\lv$ $\lambda$ couplings the final state will
consist of m leptons and $\et$; for the $\lv$ and $\bv$
$\lambda^{'}$ couplings it will consist of m leptons, n jets and $\et$
whereas $\bv$ $\lambda^{"}$ couplings give rise to final state with only jets.
The authors considered different sources of background in each case, 
chose different points in the parameter space to consider the 
chargino/neutralinos states with different gaugino/higgsino content 
and studied the process in a parton level Monte Carlo, with appropriate 
cuts on leptons and jets to reduce/remove background.  
\begin{table}[htb]
\caption{\em Showing the contributions (in fb) of different (light)
chargino and neutralino production modes to multi-lepton signals at the
NLC in the case of $\lambda$ couplings. The last column shows the SM
background.\label{tab:tab3}}
\begin{center}
\begin{tabular}{|c|l|cccc|c|c|}
\hline  
&&&&&&&\\
&Signal       
&$  \N0_1 \N0_1 $ & $ \N0_1 \N0_2$  & $ \N0_2 \N0_2$  & $\Cp_1 \Cm_1$ &  
{\bf Signal} & {\bf Bkgd.}  \\
&&&&&& fb & fb\\
\hline 
{\bf A} 
&$1\ell + \mET$ & $1.1$ & $0.4$ & $0.2$ & $1.5$   &$3.2$ & $8272.5$
\\ \cline{2-8}
& $2\ell + \mET$ & $ 14.9$ & $5.2$ & $1.8$ & $15.3$   &$ 37.2$ &$ 2347.4$ 
\\ \cline{2-8}
& $3\ell + \mET$ & 91.7 & 25.3 & 7.2 & 71.6   & 195.8 & 1.5
\\ \cline{2-8}
& $4\ell + \mET$ &  212.8 & 49.6 &13.6 & 152.8   & 428.8 & 0.4 
\\ \cline{2-8}
&$5\ell + \mET$ &  0.0 & 37.8 & 19.3 & 113.5   & 170.6 &  -
\\ \cline{2-8}
& $6\ell + \mET$ &  0.0 & 39.6 & 21.6 & 26.9   & 88.0 &  -
\\ \cline{2-8}
& $7\ell + \mET$ & 0.0 & 0.0 & 11.9 & 0.0   & 11.9 &  -
\\ \cline{2-8}
& $8\ell + \mET$ &0.0 &0.0 &8.0 &0.0   &8.0 &  -
\\ \hline
\end{tabular}
\end{center}
\end{table}
Table~\ref{tab:tab3} shows for a particular point results of the Monte 
Carlo analysis for the case of $\lambda$ couplings.
In case of these signals, particularly for the case of $\lambda^{"}$
couplings the final state involves a large number of partons with/out
leptons. Some of these partons may merge together in jet definitions,
removing the connection between the jet multiplicity and the number of initial
partons in the final state. It is pointed out that kinematic mass reconstructions
can be used to study the multijet events and identify those coming from
$\bv$ couplings. Fig.~\ref{fig5} shows the distribution in invariant mass
constructed from the hardest jet and all the other jets in the same
hemisphere, and the same for the hardest jet in the opposite
hemisphere.
The distribution shows clear peaking at $m_{\tldch_{1}^{0}}$ as well as a
sharp cutoff at $m_{\tldch_{1}^{\pm}} \approx m_{\tldch_{2}^{0}}$. Thus
kinematic distributions can be used quite effectively even for the case of
the multijet events. Effect of backgrounds on this distribution need
to be studied, however.
\subsection{Wino LSP in AMSB at \eplem\ colliders.\protect\cite{6}}
\begin{figure}[htb]
       \centerline{
      \includegraphics*[scale=0.45]{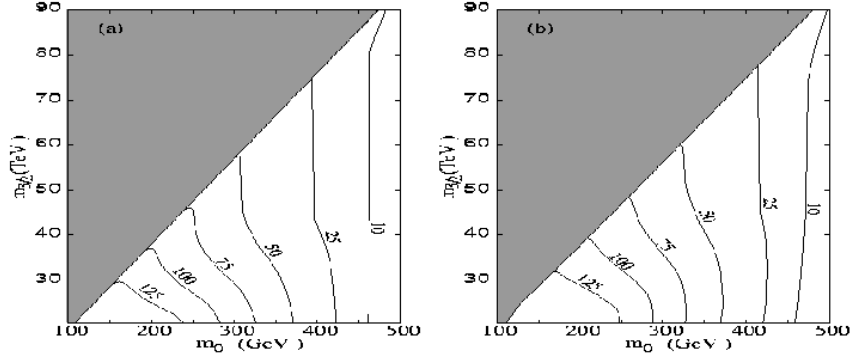}
 }
\caption{\em Regions in the $M_{3/2}-m_0$ plane along with contours of
constant cross-section expected for the signal in the AMSB scenario.
The shaded regions are ruled out by various constraints.\protect\cite{6}}
\label{fig6}
\end{figure}

In the AMSB scenario, $\tldch_{1}^{0}, \tldch_{1}^{+}$ are both almost pure
Winos and almost degenerate. The authors investigate here
$\eplem \to \tilde{e}( \to e\tldch_{1}^{0}) + \tilde{e^{*}}( \to
\nu\tldch_{1}^{+}) \longrightarrow e\tldch_{1}^{0}\nu\tldch_{1}^{+} ( \to
\tldch_{1}^{0} + \pi ) \longrightarrow e\tldch_{1}^{0}\nu\tldch_{1}^{0}\pi$.
The B.R. for $\tldch_{1}^{+} \to \tldch_{1}^{0} + \pi$ is $\sim 96-98\%$.
The authors look at the regions in $M_{3/2} - m_{0}$ plane ($m_{0}$ being
the scalar mass added to make the slepton masses nontachyonic) for a range
of $tan\beta$ values. They found that large regions, allowed by all the
currrent constraints, have large cross-sections $\sim 10-100$ fb at $\sqrt{s}$
= 1000 GeV, after kinematical cuts on the $e^{-}$ and B.R. are
included, as shown in Fig.~\ref{fig6}.
 The signal is a fast e($\mu) + \et$ and a soft $\pi$. The soft
$\pi$ can  give rise to a displaced vertex with impact parameter
resolved if c$\tau < $ 3cm. If $\tldch_{1}^{+}$ decay length is long (c$\tau
> $ 3cm), then one sees a heavily ionising track. The authors looked at the
cross-section after cuts and find that with 50 (500) $fb^{-1}$ integrated
luminosity one expects $\sim 10^{3} (10^{4})$ events at $\sqrt{s}$ = 500
(1000) GeV. It is also to be noted that this case is different from the
almost degenerate $\tldch_{1}^{+} , \tldch_{1}^{0}$ scenario in the MSSM
where the chargino/neutralino are higgsinos.\cite{Manuel}

\subsection{Squark, slepton production at \gamgam\ colliders and decays
through \rpv\ interactions.\protect\cite{7}}  
\begin{figure}[htb]
       \centerline{
      \includegraphics*[scale=0.53]{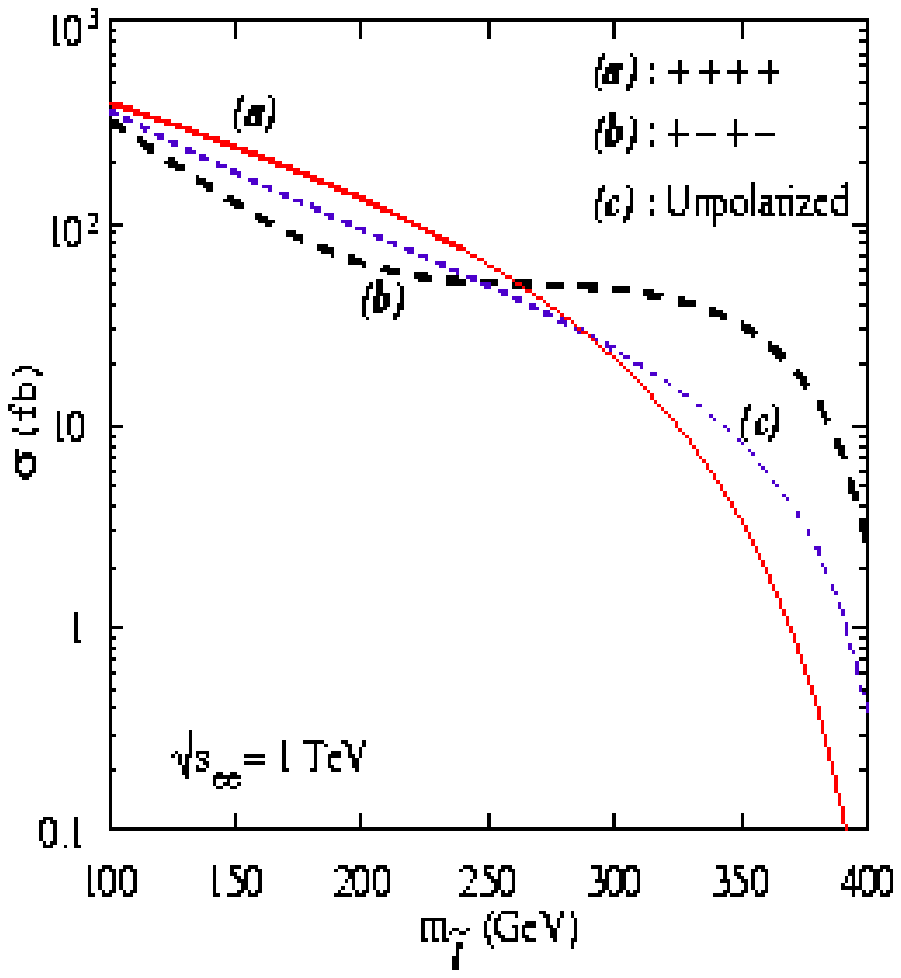}
      \includegraphics*[scale=0.50]{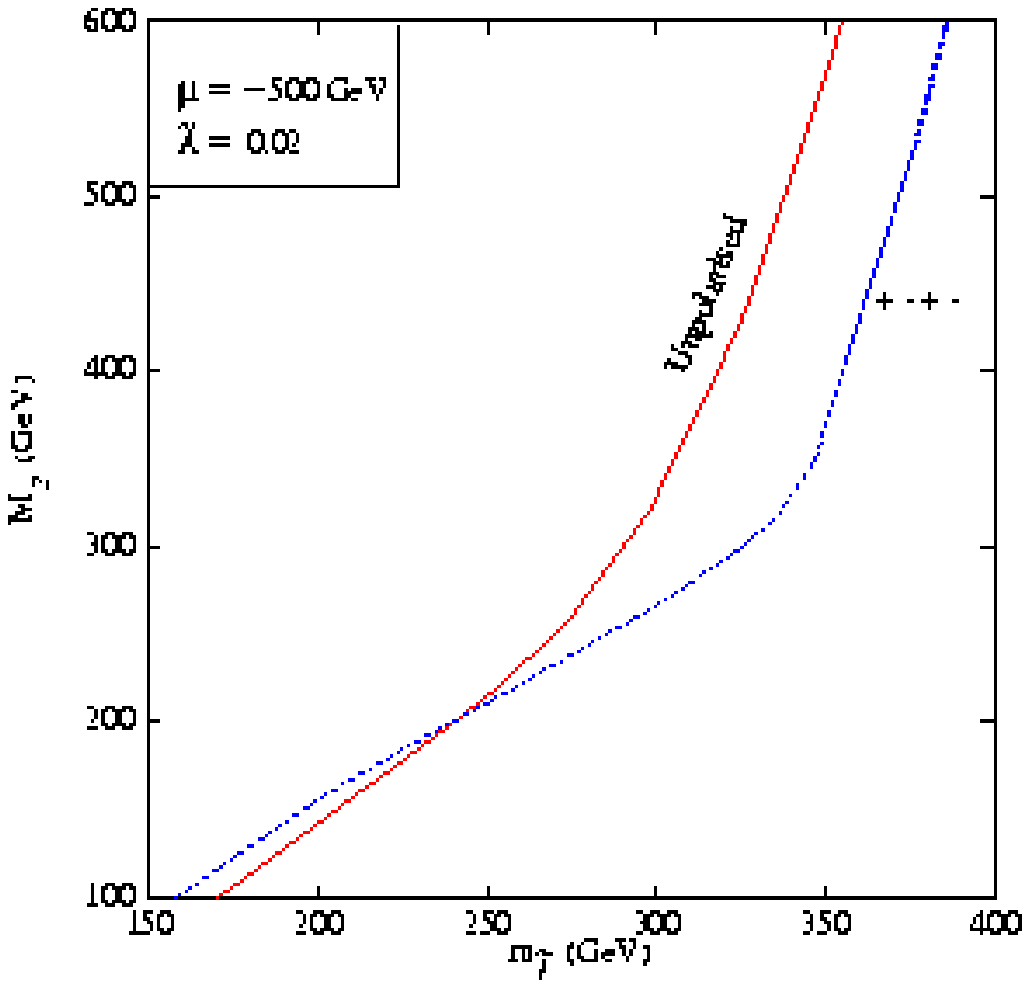}
 } 
\caption{\em Production cross-section of sfermions as a function of 
$m_{\tilde f}$(left panel), reach in $M_2,m_{\tilde f}$ plane with polarised 
and unpolarised option(right panel).\protect\cite{7}
All the parameters are as shown in the 
figure.} 
\label{fig7}
\end{figure}

The authors performed a study of $\gamgam \to \tilde{f}{\tilde{}}f^{*}$,
followed by the $R_{p}$ conserving decays $\tilde{f} \to f + \tldch_{1}^{0},
\tilde{f} \to f' + \tldch_{1}^{+}$ along with the \rpv\ 
decays $\tilde{f} \to f_{1} f_{2}$ where $f_{1} , f_{2}$ 
are SM fermions ($q_{1} q_{2}$ for $\tilde{l}$, lq
for $\tilde{q}$). One of the features worth nothing is that the 
production  cross-section of scalars can be enhanced by an appropriate
choice of polarisation.\cite{CMG} This can be seen in Fig.~\ref{fig7} where
$\sigma (\gamgam \to {\tilde{l}}^{+}{\tilde{l}}^{-} )$ has been plotted for
different polarisation combinations for $\sqrt{s_{\eplem}}$ = 1 TeV, using
the back-scattered laser spectrum. They have then calculated the
branching ratios for the \rp\ conserving as well as \rpv\ two body decays.
These of course depend on the SUSY model parameters $M_2,\mu$. 
The signal for \rpv\ decays is simply 4f
final states. The authors use kinematic cuts to reduce the background, e.g.
from heavy flavours. Further, they reconstruct lj, $j_{1}j_{2}$ invariant
masses and demand that $| M_{lj}^{(1)} - M_{lj}^{(2)} |$ $ < $ 10 GeV for
the squark signal and $| M_{ij} - M_{kl} |$ $ < $ 10 GeV for the slepton signal.
The combinatorial background is quite high for the second case. The panel on
the right in Fig.~\ref{fig7} shows the reach in $M_{2} - m_{\hat{l}}$ plane
for $\lambda'$ coupling = 0.02. The dependence on $M_{2}$ comes from the
dependence of the \rpv\ B.R. on $M_{2}$.

\section{Probing extra dimensions at $\eplem / \gamgam / e\gamma / \emem $
colliders. }

\subsection{Looking for the radion in RS model at $\eplem$ colliders.\protect\cite{10}}
\begin{figure}[htb]
       \centerline{
      \includegraphics*[scale=0.30]{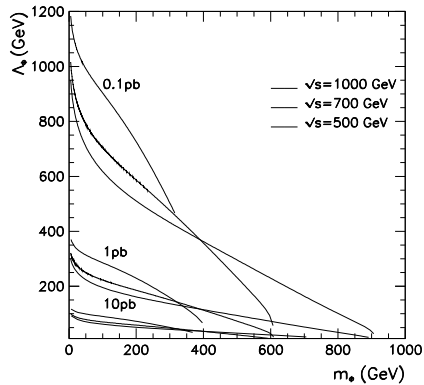}
  }
 \caption{\em Regions in the $\Lambda_\phi$ - $m_\phi$ plane along with 
contours of constant cross-sections for the production of 
$\phi$.\protect\cite{10}}
\label{fig8}
\end{figure}

These authors have looked at the phenomenology of the radion field $\phi$
which stabilizes the RS scenario.\cite{RS} This field can be lighter than the KK
excitations and its couplings to the SM particles are determined by
general co-variance in four dimensional space-time. There are two parameters:
$m_{\phi}$ and a scale $\Lambda_{\phi} \sim 0(v)$ where $v$ is the
vacuum expectation value of the Higgs field.
 The authors have computed two body decay modes of $\phi$ whose couplings to
fermions are $\sim g^{SM}_{hf\bar f} {v}{\Lambda \phi}$. Regions
for low $\Lambda_{\phi}-m_{\phi}$ are ruled out by L3 bound on $m_{h}$ and
the requirement of perturbative unitarity\cite{Mahanta} of the couplings of
$\phi$. Cross-sections for production of $\phi$ at $\eplem$
colliders have been computed and the region in $m_{\phi}-\Lambda_{\phi}$
plane that can be probed has been identified. This is shown in
Fig.~\ref{fig8} as contours of constant cross-section in the plane.

\subsection{Indirect effects of large extra dimensions.\protect\cite{11,11p,12,13}}
These authors have looked at essentially the ADD model\cite{ADD} and studied
the indirect effects on $t\bar t$, dijet production at $\gamgam$ colliders
and graviton production in $e\gamma$ collisions. They have also studied the
indirect effects of the gravitons in the $\emem$ collisions in the
RS model.\cite{RS}  In the study of dijet/$t\bar t$ production\cite{11,11p}, 
the authors have used the
idealized backscattered laser spectrum and the total integrated luminosity
of 100 $fb^{-1}$. They observe that the reach of these colliders for the
scale $M_{s}$ can be increased substantially by an effective use of
polarisation. They have obtained the possible bounds by looking only at the
total $t\bar t$ cross-sections. 
\begin{table}[htb]
\caption{\em Limits on the scale $M_s$ that can be reached\protect\cite{11p} in 
top production at \gamgam\ colliders for polarised leptons and lasers in case of 
ADD model.\protect\cite{ADD}\label{tab:tab4}}
\begin{center}
 \begin{tabular}{|c|c|c|}
 \hline
$ (\lambda_{e1} \lambda_{e2} \lambda_{l1} \lambda_{l2})$ &$\sqrt{s}$ GeV 
& $M_{s}$ (TeV) \\ \hline
& 500 &1.95 \\
(+ - - -) & 1000&4.6 \\ 
 &1500 & 6.0 \\ \hline
& 500&2.5\\
(+ - - +) & 1000 & 4.8 \\
&1500&6.4\\ \hline
\end{tabular}
\end {center}
\end{table}
Table~\ref{tab:tab4} shows their results. These
bounds have been obtained by using only the statistical errors. The analysis
can be improved by using distributions in kinematic variables as well as by
considering the systematic errors. In dijet production, e.g. the `resolved'
photon contribution\cite{zpc1} could be nontrivial, which has not been 
considered in the analysis.

Direct study of graviton production in $e\gamma \to eG^{(n)}$ can also be
used to probe the `large' extra dimensions.\cite{12} The signature of such graviton
production will be an isolated $e^{-}$ and missing energy. The backgrounds
are $e\gamma \to eZ \to e\nu \bar \nu$ and $e\gamma \to \nu w \to \nu \nu
e$. These backgrounds had been evaluated in a study of $e\gamma \to
\tilde{e} \tldch_{1}^{0} \to e + \tldch_{1}^{0} + \tldch_{1}^{0}$.\cite{8}
Using idealized backscattered laser spectrum, L = $\int {\cal L} dt = 100 \:
fb^{-1}$, demanding $\sigma \sqrt{L}/\sqrt{B} > 5$ and putting cuts
on $P_{T}^{l}, |Y_{e}|$ to remove the background, the reach for $M_{s}$ in
the ADD model\cite{ADD} is between 4-2 TeV for n=2 to 6. The use of $e^{-}$
polarization to reduce eW background is absolutely essential. Using
polarized lasers might improve the reach even further, but it has not been
studied yet.

Indirect effects in $\emem$ collisions in the RS model 
also provide a good reach
for the graviton mass $M_{1}$. The analysis looks at $\emem \to \emem$ and
puts cuts of $\theta_{e} > 10^{0}, P_{T}^{l} > 10$ GeV for the detected
electrons. They state their results in terms of $c_{0} = \frac{1}{8\pi}
\frac{{\cal K}}{\overline {M_{p}}}$ where $\overline {M_{p}}$ is 
the reduced  Planck mass 
and ${\cal K}$ is the extra mass scale in the model.\cite{RS} 
\begin{table}[htb]
\caption{\em Limits on mass $M_1$ of the first graviton excitation in the 
RS model\protect\cite{RS} at an $e \gamma$ collider for different beam 
energies, luminosities and model parameter $c_0({\cal K})$.\label{tab:tab5}}
\begin{center}
 \begin{tabular}{|c|c|c|c|}
 \hline
 $\sqrt{s}$ & $\int L dt $ ( fb $^{-1}$) & $c_{0}$ & $M_{1}$ (TeV) \\ \hline
 500 & 500  & 0.01 & 1.3 \\ \cline{2-4}
 & 500  & 0.1 & 4 \\ \hline 
 1000 & 500  & 0.01 & 2.4 \\ \cline{2-4}
& 100  & 0.1 & 6.4 \\ \hline 
 \end{tabular}
\end {center}
\end{table}
Table~\ref{tab:tab5} gives the reach\cite{13} for mass $M_{1}$ of 
the first graviton excitation for different beam energies, luminosities 
and values of model parameter $c_0({\cal K})$. Here use of polarisation 
improves the reach. This analysis
has used the information on the distribution of $e^{-}$ and 
assumed a modest polarisation of 80\%. The estimates of error, however, 
are only statistical ones.

\section*{Acknowledgments}
I wish to thank the organisers of the APPC 2000 and III ACFA LC Workshop
for organising an excellent meeting and providing a very nice atmosphere
for discussions.

\end{document}